\documentclass{aa}  
\usepackage{graphicx,hyperref,url} % Required for inserting images
\newcommand{\xmm}{{\em XMM-Newton}}
\newcommand{\lnine}{{L\,98-59}}
\newcommand{\fxu}{erg s$^{-1}$ cm$^{-2}$}
\newcommand{\nhu}{cm$^{-2}$}
\newcommand{\lxu}{erg s$^{-1}$}
\newcommand{\logrhk}{$\rm log\,R^{\prime}_\mathrm{HK}$}

\newcommand{\mearth}{M$_{\oplus}$}

\usepackage{txfonts}
\titlerunning{X-rays from \lnine}
\authorrunning{Pillitteri et al.}

\begin{document} 
\title{Long-term X-ray variability of the multiple-planet host L~98-59: Hints of an activity cycle.}

\author{I. Pillitteri\inst{1} \and
        S. Bellotti\inst{2,3} \and
        S. Benatti\inst{1} \and
        S. Boro Saikia\inst{4} \and 
        A. Garc\'ia Mu\~noz\inst{5} \and
        K. G. Kislyakova\inst{4} \and
        A. Maggio\inst{1}  \and
        G. Micela\inst{1}  \and
        K. Vida\inst{6,7}     \and
        A. A. Vidotto\inst{2} 
          }

\institute{ 
         {INAF-Osservatorio Astronomico di Palermo, Piazza del Parlamento 1, 90134 Palermo, Italy\\ \email{ignazio.pillitteri@inaf.it} }
        \and 
         {Leiden Observatory, Leiden University,
            PO Box 9513, 2300 RA Leiden, The Netherlands}
        \and
            {Institut de Recherche en Astrophysique et Plan\'etologie,
            Universit\'e de Toulouse, CNRS, IRAP/UMR 5277,
            14 avenue Edouard Belin, F-31400, Toulouse, France }
        \and
        {Department of Astrophysics, University of Vienna, 
        Türkenschanz Strasse 17, 1180, Vienna, Austria} 
        \and 
        {Universit\'e Paris-Saclay, Universit\'e Paris Cit\'e, CEA, CNRS, AIM, Gif-sur-Yvette, 91191, France}     
         \and
        {Konkoly Observatory, HUN-REN Research Centre for Astronomy and Earth Sciences, Konkoly Thege Miklós út 15-17., H-1121, Budapest, Hungary}
        \and 
        {HUN-REN CSFK, MTA Centre of Excellence, Budapest, Konkoly Thege Miklós út 15-17., H-1121, Budapest, Hungary}
}
   \date{Received; accepted }

\abstract
{ 
% \LEt{***General notes: in title, i.e.,  The star L 98-59 hosts more than one planet, yes? (needs the hyphen and singular)\\a) I edited to US English convention. \\b) A\&A uses the past tense to describe the specific steps used in a paper and the present tense to describe general methods and recent findings. Please make sure this is followed throughout. For details, see Sect. 6 of the Language Guide: https://www.aanda.org/ for-authors/language-editing/6-verb-tenses \\c) Instrument and program names are introduced (when appropriate) at first use in the main text. All other abbreviations and acronyms are introduced at first use, once in the Abstract and again in the main text. All abbreviations should be used consistently. Please check and amend as necessary throughout.}\\\\
High-energy irradiation in X-rays and UV (XUV) can transform the planetary 
atmospheres through  photoevaporation and photochemistry. 
This is more crucial for M stars, whose habitable zones for Earth-like planets are located within a few percent of an AU.
Transiting exoplanets around M dwarfs offer the opportunity to study their 
characteristics and habitability conditions.
\lnine\ is an M3 dwarf {hosting six Earth-like planets, with two of them in the 
habitable zone of the star}. 
X-ray observations made in 2020 and 2021  detected significant flares above a 
quiescent luminosity of $4-10\times10^{26}$ \lxu.
We present the results from two short \xmm\ observations of \lnine, {which are part of a  
monitoring survey to detect long-term X-ray variability and activity cycles.} 
In October 2024 the X-ray quiescent luminosity of the star was $\sim5.9\times10^{25}$ \lxu, 
and it was about $6.3\times10^{26}$ \lxu in February 2025. 
We speculate that in  late 2024 the star had a minimum of activity; in
2021 the star was near a maximum {of an activity cycle}, and in 2025 it was at the middle 
of the cycle. {We suggest a coarse estimate of the period of $\approx2$ years} and a  
peak-to-peak amplitude of about $\simeq10$, which is the highest among the stars with a known 
X-ray cycle other than the Sun.
{We also infer that even the outer planet in the habitable zone, \lnine\~f, is exposed
to an X-ray dose between 100 and 1600 times the X-ray irradiation of the Earth in the XMM band.}
}


   \maketitle

\section{Introduction}
The discovery of the solar magnetic activity cycle—revealed by periodic variations in sunspot 
numbers—was one of the earliest astrophysical milestones (Schwabe 1844). Similar chromospheric 
cycles have since been detected in many stars through long-term Ca II H\&K monitoring (e.g., 
Baliunas et al. 1995), with periods ranging from months to decades. However, the high-energy 
counterpart of these cycles, in X-rays, remains poorly explored. To date, only seven stars are known to 
host X-ray cycles (e.g., Robrade \& Schmitt 2016; Coffaro et al. 2020, 2022).

The scarcity of detections reflects the challenges of X-ray monitoring, but also represents a 
major gap in our understanding. X-ray emission traces the hot corona and dominates the high-
energy irradiation of exoplanets. Its variability on cycle timescales can strongly influence 
atmospheric evaporation and photochemistry, processes that ultimately determine the fate of 
planetary atmospheres. However, whether X-ray cycles follow the same behavior as chromospheric 
cycles, and how their amplitude and timescales depend on stellar age and type, remain open 
questions.
The aim of our program is to close this gap with the first systematic, multi-year monitoring of a 
well-characterized stellar sample. By focusing on transiting-planet hosts in the Ariel 
Mission Reference Sample, the results can have a relevant impact on exoplanet 
atmosphere studies.

Photoevaporation and photochemistry processes among exoplanets are determined by
the high-energy stellar irradiation in X-ray and EUV (XUV) bands,  the stellar activity 
and the frequency of XUV flares, and Coronal Mass Ejections (CMEs).
The effects of long timescale modulation of XUV irradiation are still to
be understood due to the sparse knowledge of stellar cycles in the XUV band \citep{Jeffers2023}.

While the search for chromospheric activity cycles in stars   received a strong impulse
by the monitoring operated at the Mount Wilson Observatory between  1966 and 1995\footnote{\url{https://dataverse.harvard.edu/dataverse/mwo_hk_project}} \citep{Vaughan1978,Wilson1981}, 
in X-rays there are very few cycles detected among FGK stars because of the sparse observations.
X-ray cycles have been found in Alpha
Cen A \& B \citep{Robrade2016}, Prox Cen \citep{Wargelin2024}, 
61 Cyg A \& B \citep{Hempelmann2006,Robrade2012}, HD 81809 \citep{Favata2008,Orlando2017},
$\iota$ Hor (\citealp{Sanz-Forcada2013}), $\epsilon$ Eri 
(\citealp{Coffaro2020}), and Kepler 63 (\citealp{Coffaro2022}). 
From this small sample, it emerges that X-ray cycles have shorter periods and 
smaller peak-to-peak amplitudes in younger stars \citep{Favata2008}, and this behavior
could be  ascribed to a larger filling factor of active regions in the coronae with higher 
levels of activity.  In 61~Cyg~A the periods of chromospheric and
X-ray cycles are similar to those of the Sun \citep{Boro-Saikia2016, Boro-Saikia2018}. 
A short cycle of 122 days has been detected in $\tau$ Bootis in the chromospheric and optical bands 
\citep{Mittag2017, Jeffers2018} in rough agreement with a similar cycle found in X-rays. 
Also worth noting is the 1.6-year chromospheric cycle found in $\iota$ Hor 
(age t$\sim$0.6 Gyr), which also shows  a beating pattern % \LEt{***an alternating pattern?} 
\citep{Alvarado2018}. 
Peculiarly,  in  $\epsilon$ Eri the X-ray and chromospheric cycles did not follow the same trend \citep{Coffaro2020},  
suggesting that X-ray cycles can behave differently from chromospheric cycles for reasons that are not yet understood.

We {embarked on} a monitoring survey with \xmm\ of a selected list of targets for the 
ESA Ariel mission \citep{Tinetti2018}. The aim was to discover more X-ray cycles among planet hosts and
to determine whether these cycles can have any effects on the atmospheres of their planets. % \LEt{***have I interpreted correctly?}
Knowing the X-ray cycle could help to plan the optical observations of planetary atmospheres 
during the low-activity phases, thus minimizing  the effects related to the stellar activity.
Similar efforts for Ariel have also been conducted  from ground-based  spectropolarimetric 
observations \citep{Bellotti2024}.
Here we present the first two observations from our program targeting \lnine\ (TOI-175), 
which is an M3 dwarf at  10.6 pc from the Sun \citep{Cloutier2019}.
The system of \lnine\ {is composed of} at least six planets with masses  
between 0.4 \mearth\  and 3 \mearth;   planets {\it e and f} are in the habitable 
zone, and planet {\it d} likely possesses water and a gaseous envelope \citep{Demangeon2021}.
\lnine\ has also been  a target of JWST transmission spectroscopy;  
sulfur was detected in the atmosphere of the 1.6 \mearth\ planet~{\it d} \citep{Gressier2024}.  
Transiting planets around M dwarfs similar to 
% \LEt{***A\&A discourages the use of "like" since it can lead to ambiguity; "such as" and "similar to" are preferred alternatives depending on the context. Please check throughout that I have interpreted correctly.} 
\lnine\  are particularly appealing since their putative habitable zones are within 
a few tenths of AU, corresponding to periods of a few days. They are thus suitable 
for transmission spectroscopy and atmospheric composition retrieval. 
\lnine\ thus constitutes  an important system for the study of the habitability of the 
extrasolar planets around M stars in the solar neighborhood. 
Together with similar nearby M dwarfs with packed planetary systems,  for example  Trappist-1 and LHS~1140, 
\lnine\ is thus a prime target for the characterization of Earth-like planets with present
and future observatories such as JWST and Ariel.

The structure of the paper is the following. Section \ref{obs} describes the 
\xmm\ observations and their analysis. In Section \ref{results}  we present
our results. In Section \ref{discuss} we discuss the results and give our conclusions.

%--------------------------------------------------------------------
\section{Observations} \label{obs}
\lnine\ was observed with \xmm\ {in three previous epochs} (see Table
\ref{xmmlog}).
{In those observations the quiescent luminosity was about $4-10\times10^{26}$ \lxu\ 
in the band $0.3-10$ keV. 
The star  showed two prominent flares (cf. Fig. \ref{fig:lc}; see \citealp{Behr2023}), 
which is a typical signature of X-ray activity in M stars.}

Here we present the results of two further observations targeting \lnine, 
together with a re-analysis of the archival observations in a homogeneous way 
(see Sect. \ref{xmmsect}).
We also made use of archival optical spectra of \lnine\ to infer the chromospheric activity 
index \logrhk\ (see Sect. \ref{harpssect}.

\begin{figure}
   \centering
   \resizebox{0.5\textwidth}{!}{
   \includegraphics[]{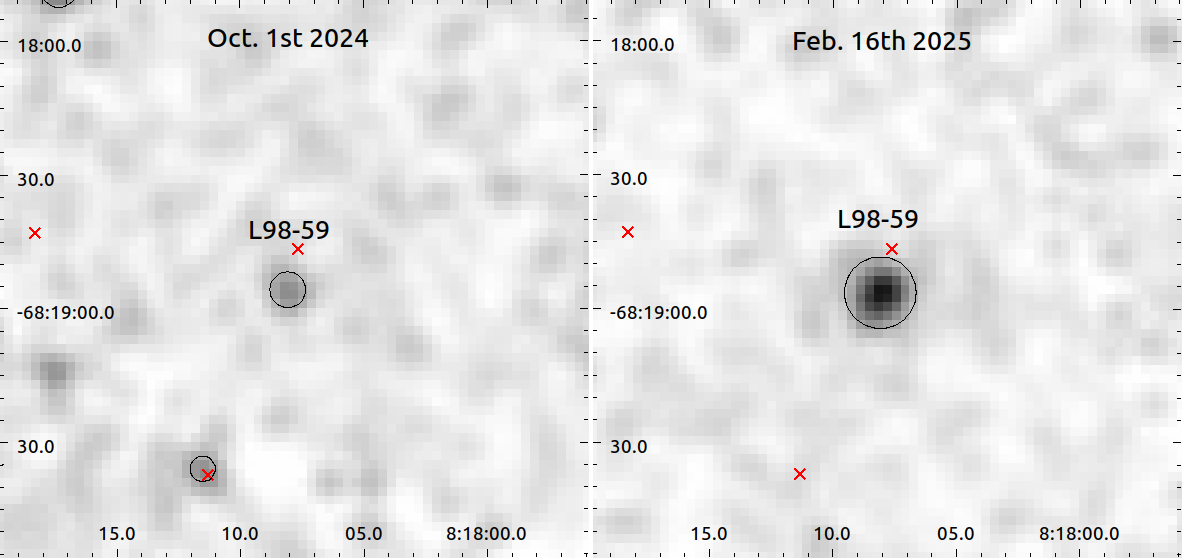}
   }
   \caption{Comparison of X-ray rate images. % \LEt{***Perhaps: Comparison of X-ray rates.} 
   The red crosses mark the positions of the objects in SIMBAD. \lnine\ has high proper motions toward 
   the bottom left of the image and its current position is coincident with the 
   centroid of the X-ray emission.
   The images are on the same pixel and color intensity scales (2"/pix, 
   maximum 2.5 counts per pixel)  and are smoothed with a Gaussian kernel with $\sigma=1.5$ pixels. 
              \label{fig:rate-comp}}
\end{figure}

\section{Results}
\label{results}
\begin{table*}[]
\caption{Count rates, detection significance, unabsorbed X-ray flux and luminosity, and hardness ratio (HR) 
of \lnine\ in each of the \xmm\ observations  during the quiescent intervals.  
}
\begin{center}
\resizebox{0.90\textwidth}{!}{
\begin{tabular}{lcccccccc} \\\hline\hline
ObsID   & Year    & Exp. time & Signif.  & SAS\_ML & Rate & f$_X$ &  L$_X$ & HR\\
          &         &  (ks)      & $\sigma$  &  & 10$^{-3}$ ct s$^{-1}$ & 10$^{-14}$ \fxu    & 10$^{26}$ \lxu \\ \hline   
0863400601 & 2020    &  27.2 & 15.6  &  100.16 & 3.93$\pm$0.53 & 3.06$\pm$0.41 & 4.13$\pm$0.6  & -0.12 (-0.60 -- 0.32)  \\  
0871800201 & 2021   &  11.6 & 17.3  &  159.25 & 9.87$\pm$1.21 & 7.66$\pm$0.80 & 10.34$\pm$1.1 & 0.14 (-0.20 -- 0.46) \\
0871800301 & 2021   &  10.2 & 12.1  &   65.9 & 7.78$\pm$1.26 & 6.05$\pm$0.77 & 8.16$\pm$1.05 & 0.08 (-0.39 -- 0.50)\\ 
(087180-Sum& 2021 &  21.8 & 20.3  &   --  & 7.84$\pm$0.82 & 6.10$\pm$0.98 & 8.23$\pm$1.32 &  0.11 (-0.15 $\div$ 0.38) ) \\
0940540301 & 2024    &  34.0 &  5.3  &  11.44 & 0.56$\pm$0.16 & 0.44$\pm$0.13 & 0.59$\pm$0.17 & -0.27 (-1.00 -- 0.53) \\ 
0940540401 & 2025    &  15.6 &  13.3 &  110.48 & 6.0$\pm$0.9 &  4.71$\pm$0.67 & 6.33$\pm$0.9 & -0.04 (-0.37 -- 0.261)  \\ \hline
\end{tabular}
}
\end{center}

\tablefoot{ Exposure time is the sum of MOS and pn exposure times in quiescent intervals. 
Significance is in units of $\sigma$ of local background.
SAS\_ML is the maximum likelihood value of the source detection performed with SAS.
Count rates are scaled to the MOS effective area. 
Unabsorbed fluxes in 0.3--10 keV band were determined assuming a 1T APEC thermal model with solar abundances, $kT = 0.17$ keV and $N_H =10^{18}$ \nhu.
Luminosities are calculated assuming a distance of 10.6 pc.
HR calculated from pn counts in the bands $S=0.3-0.7$ keV  (soft) and $H=0.7-8.0$ keV (hard).
We report the measurements from the sum of the observations 0871800201 and 0871800301.}
\label{tab:rate_lx}
\end{table*}
\lnine\ was detected in both our \xmm\ observations in October 2024 and February 2025,
but with different significance over the local background mean level. In particular 
the  significance was about 5.3 $\sigma$ in October 2024, 
while the significance of the source  was about 13.3 $\sigma$ in February 2025. 
In previous observations it was detected with a significance level comprised 
between 12.1 and 20.3 $\sigma$. 
The flux and luminosity of \lnine\ in 2024 was a factor of five to ten lower than the 
corresponding quiescent rate recorded in the other observations. 
If we also consider  the peaks of the flares recorded in 2021 \citep{Behr2023}
the difference with the lowest quiescent emission is about a factor 
100. 
The hardness ratio obtained from {\it pn} %\LEt{***abbreviation to introduce here at first use.} 
counts in the bands $0.3-0.7$ keV and $0.7-8.0$ 
keV shows a quite soft spectrum; there is a shallow positive trend of the spectral 
hardness with the flux, but its significance is affected by the low count statistics. 
In general, however,  an intrinsically harder spectrum is  expected for higher 
flux emitted from hotter plasma. 

A low level of activity and the X-ray emission from the corona and its soft spectrum 
inferred from the hardness ratio supports a very old age for \lnine. 
The decrease in  flux across the X-ray observations is likely associated with 
an overall decrease in the emission measure  of the plasma
in terms of volume of emitting plasma and number of coronal active regions. 
This means that in 2024 the corona of \lnine\  was devoid of significant active 
regions, similarly to the Sun at its minimum of activity.
The star shows some degree of variability in X-rays even at its lowest level of 
emission, but owing to the low count statistics the variability is within 3$\sigma$ of the 
mean count rate  (Fig. \ref{fig:lc}). 

\section{Discussion and conclusions}
\label{discuss}
\begin{figure}
    \resizebox{\columnwidth}{!}{
    \includegraphics[width=0.90\linewidth]{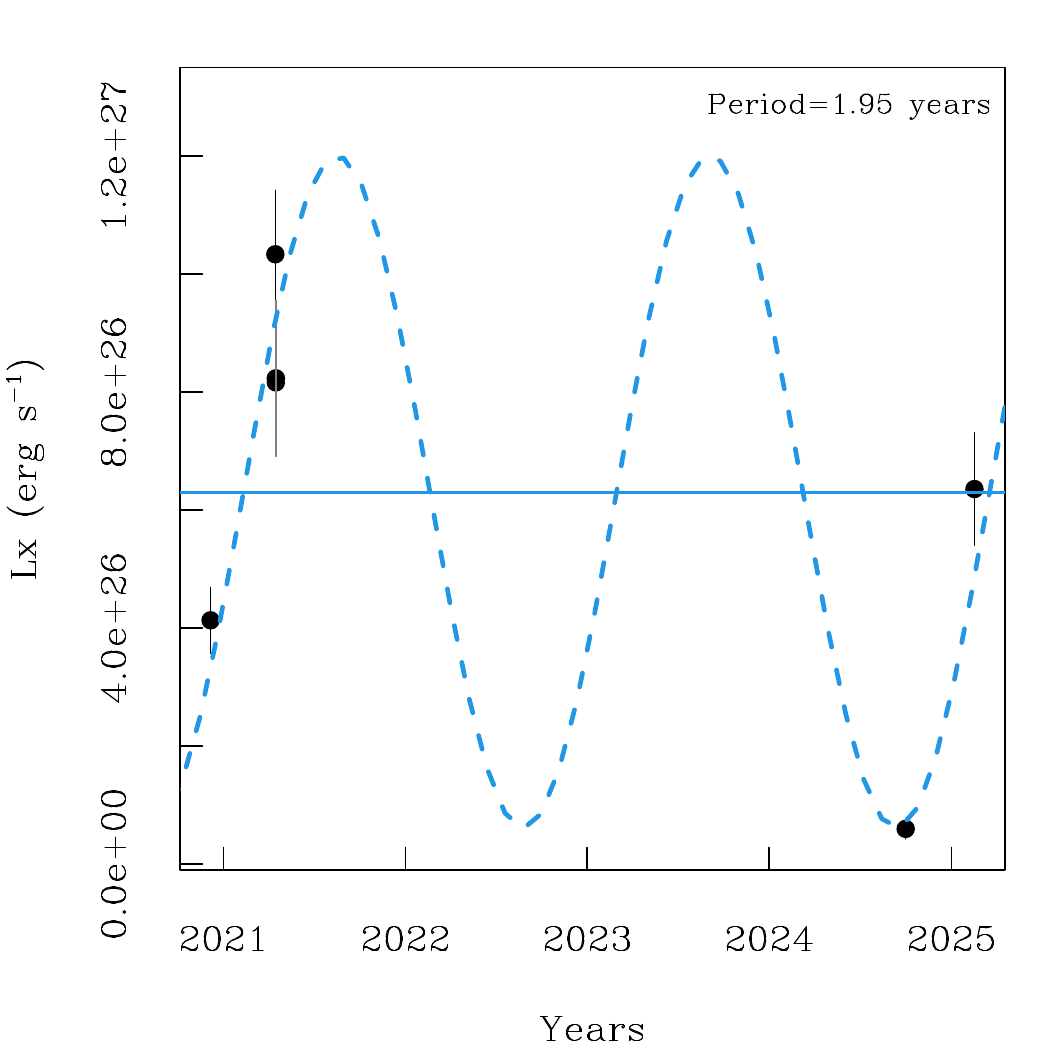}}
     \caption{X-ray luminosity of \lnine\ in the band 0.3-10 keV vs. time. 
     A cycle of about 1.95 years is shown as a sinusoidal dashed line. 
     We made a rough estimate of the period by assuming that the minimum of the sinusoid is 
    the value recorded in October 2024 and that its mean is equal to the value measured in February 2025. 
 % \LEt{***estimate? guess sounds random} 
    \label{fig:lx}}
\end{figure}

In a 10 ks \xmm\ observation of \lnine\ obtained in October 2024  we detected an X-ray flux 
 that was about ten times lower than the flux measured in the other observations, with a significance of more
than 3 $\sigma$.
In February 2025 the flux was between the values recorded in 2020 and in 2021 (Fig. \ref{fig:lx}). 

We speculate that an activity cycle in X-rays could be at work in \lnine\ with a period of 
about 2 years (Fig. \ref{fig:lx}), although we are aware that more monitoring is required
to firmly detect such a cycle.   
The star was near its minimum of activity during late 2024 and had a maximum of activity
presumably in mid-2021 (and another unmeasured peak in 2023) plus intermediate states recorded 
in December 2020 and February 2025.  
Near the maximum the star was characterized by frequent flaring activity
together with a higher basal X-ray emission (see \citealp{Behr2023}). 

\lnine\ was  detected {by eROSITA in the eRASS1} catalog \citep{Merloni2024} with a log L$_X = 26.48$ and
a ratio $\log \mathrm L_X / \mathrm L_\mathrm{bol} = -5.14$. The luminosity from eROSITA 
is similar to the \xmm\ luminosity measured in 2020.

\begin{figure}
    \centering \resizebox{\columnwidth}{!}{
    \includegraphics[]{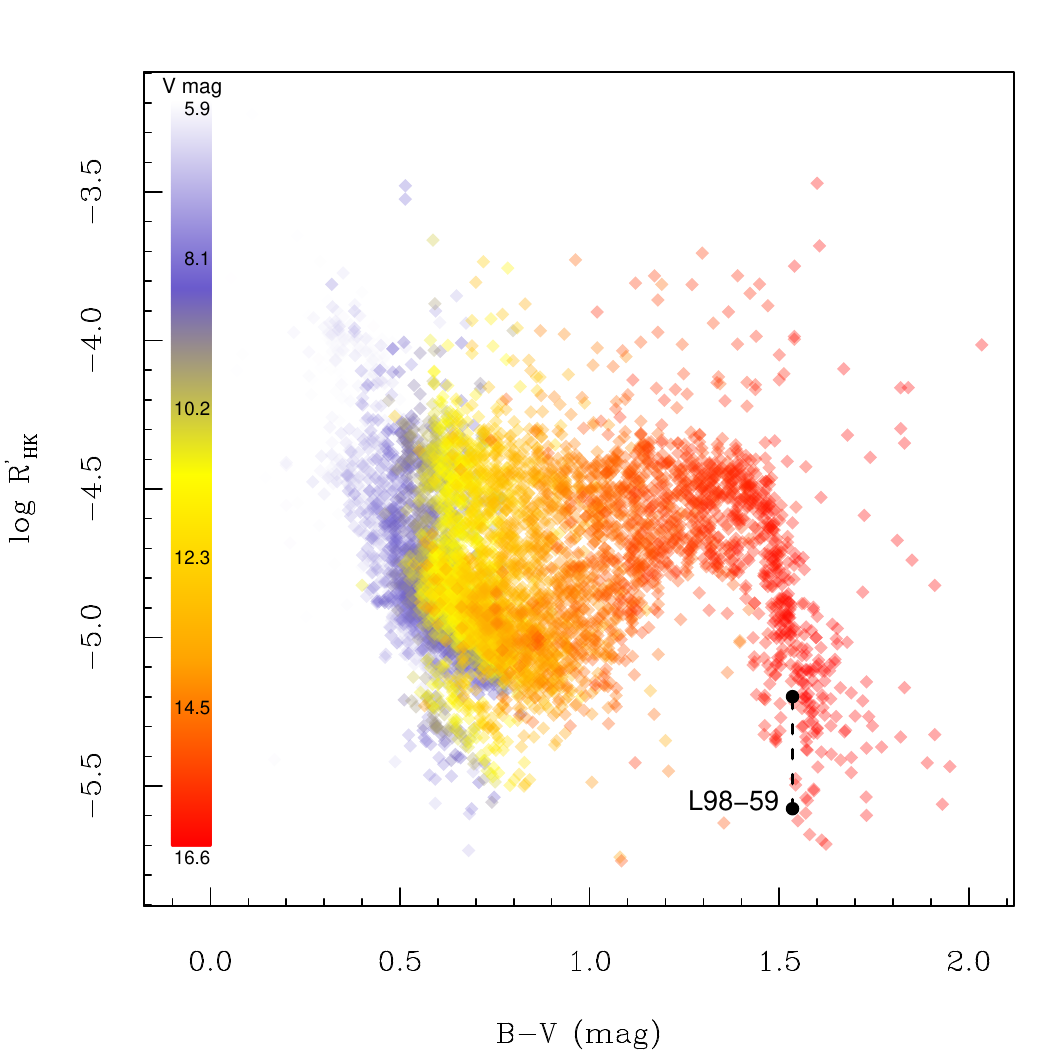} }
    \caption{Index \logrhk\ {for F to M dwarfs} from \citet{Boro-Saikia2018} and the range of \logrhk\ observed in \lnine. }
    \label{fig:logRhksample}
\end{figure}

\citet{Boro-Saikia2018} reported measurements of \logrhk\ for 4454 stars along the
main sequence from F to M dwarfs. We   compare the 5\%-95\% quantile range variation
of \lnine\ with their sample in Fig. \ref{fig:logRhksample}. At \logrhk=-5.6, 
which corresponds to the 5\% quantile, the star is one of the most inactive in the sample
of M dwarfs, and when \logrhk\ reaches the 95\% quantile at $\sim-5.2$ the star appears
more similar to the other M dwarfs. 

Overall, the star looks very old and inactive; an estimate of the age based on its rotation 
period and \logrhk\  points to $t\geq5$ Gyr \citep{Boudreaux2022,Engle2023}.
Compared to other mid-M stars with a similar rotation period of 80 days, \lnine\ {lies}
at the bottom of the X-ray luminosity distribution when it is at the minimum of activity 
\citep[cf. Fig. 2 in][]{Magaudda2020}. 

\lnine\ would be the second M dwarf with a potential activity cycle in X-rays together
with Prox Cen \citep{Wargelin2017, Wargelin2024}. 
Prox Cen and \lnine\ have similar rotation periods (89 days vs. 80 days, \citealp{Medina2020}), 
the spectral type of Prox Cen  is later than that of \lnine. However, while the X-ray cycle in 
Prox Cen has an amplitude of a factor of 1.5, in \lnine\ the amplitude could be larger as it
does show a peak-to-peak amplitude of a factor of about ten.
The other known stars with X-ray cycles are either G or K stars, often members of 
binary systems.

\citet[Fig. 20]{Coffaro2022} compared the X-ray surface fluxes and ages of stars 
with cycles with the amplitude of their X-ray cycles  ($L_{X,max} / L_{X,min}$) 
noting that higher fluxes due to young age and higher activity correspond to cycles with lower 
amplitudes or absent cycles. 
\lnine\ is consistent with the empirical relationship obtained by
\citet{Coffaro2022} only when the star is at its lowest activity level, with a surface X-ray 
flux of about  10$^4$ \fxu\
and for an age of $t\geq5$ Gyr, while it appears a strong outlier when it is at the maximum
of activity bearing  an X-ray surface flux one order of magnitude higher.

In the absence of an activity cycle, the sharp drop in luminosity could be attributed to
a corona with a more active  X-ray bright side and a less active X-ray dark side. 
However, there is a low probability that such a dark side would have been 
spotted  in only one short exposure out of five observations. 
The interpretation that there is an activity cycle
appears the most viable; 
% \LEt{***The interpretation of an a. cycle? The interpretation that it is an a. cycle?} 
however more monitoring  is needed to determine the nature of the X-ray variability 
of \lnine\ that we observed on a timescale of a few months.

% \LEt{***One-sentence paragraphs should not be used. Please check throughout, and include single sentences in the previous or following paragraph, as appropriate, or rephrase (i.e. add at least one more sentence).}
The X-ray hardness ratio of \lnine\ across the observations has a slight
trend, but overall it is the feature of a steady soft spectrum during 
the quiescent phases.
{The low flux recorded in 2024 could be due to a low plasma emission measure in   
an X-ray dark corona devoid of active regions. 
In this configuration, a steady stellar wind would originate from an 
open magnetic field line configuration. 
The maximum of activity instead would be characterized by frequent flares (Fig. \ref{fig:lc}) 
and perhaps CMEs occurring during the most powerful events.  
% \LEt{***see note 2c} 
In this phase of the cycle the corona would be composed of X-ray bright active regions with a closed 
magnetic field.} 
These two different configurations can have an impact on 
the dynamics and chemistry of the atmospheres of the planets around \lnine. 
The steady stellar wind could erode the planetary atmospheres on timescales of several hundreds 
of megayears or gigayears,
while flares and CMEs can both produce effects on short timescales  
related to heating, evaporation, high-energy particle interaction, and photochemistry. 
% \LEt{Units should be written out when they aren't directly accompanied by numerals.} 

We calculated the irradiation of the planets in the XUV band $5-920$ \AA\ 
for different levels of X-ray luminosity of the host star {(Table \ref{tab:planet_values})}. 
The EUV luminosity in $100-920$ \AA\ was estimated {to be about} ten times the X-ray
luminosity in the ROSAT band \citep[see Fig. 3 in ][]{Sanz-Forcada2025}.
{The XUV flux varies between  $\sim34$ and $18,200$ \fxu\ across the six planets.}
For reference, the Earth's  X-ray {irradiation}  is about 0.35 \fxu\ in the same band.
Even the outer planet {\it f} orbiting in the habitable zone of \lnine\ is
exposed to an XUV dose between 100 and 1600 times the X-ray irradiation of the Earth.
{Such irradiation can produce several effects}, for example  photodissociation, photoionization, photochemical reactions, deposits of heating due to scattered secondary electrons, and evaporation 
% \LEt{***don't forget: A\&A uses the serial comma (Oxford comma) between three or more items in a list: a, b, and c; x, y, or z. Please check and amend throughout as appropriate.  Have I grouped correctly?} 
\citep[see][]{Cecchi-Pestellini2006}. In extreme cases planets can lose their
atmospheres due to intense {irradiation} acting for several gigayears  
\citep{Lammer03,Fromont2024,VanLooveren2025}.
Highly X-ray irradiated upper atmospheres can exhibit signatures of molecular lines in the
near-IR, which can be probed by NASA JWST now, and will be probed by ESA Ariel in the {near} future 
\citep{Cecchi-Pestellini2009,Locci2024}. 

\begin{table}[t]
\caption{XUV (5-920 \AA) fluxes in log scale received from planets b–e at three different levels of 
XUV luminosity. }
\begin{center}
\resizebox{0.49\textwidth}{!}{
\begin{tabular}{l|cccccc}
\hline \hline
 planet & b & c & d & e & f & g \\
d (AU) &  0.0219  & 0.0304 &  0.0486 & 0.0717 & 0.1053 & 0.0188 \\
\hline
$\log$ L$_\mathrm{XUV}$ (\lxu) & \multicolumn{6}{c}{ $\log f_\mathrm{XUV}$ (\fxu) } \\
27.03 & 2.90 & 2.62 & 2.21 & 1.87  & 1.54 & 3.03 \\
28.06 & 3.93 & 3.64 & 3.24 & 2.90  & 2.57 & 4.06 \\
28.26 & 4.13 & 3.84 & 3.44 & 3.10  & 2.77 & 4.26 \\
\hline
\end{tabular}
}
\tablefoot{The separation adopted for flux calculation is listed under each planet.}
\end{center}
\label{tab:planet_values}
\end{table}

\begin{acknowledgements}
IP acknowledges support from Bando per il Finanziamento della Ricerca Fondamentale 2024 
dell’Istituto Nazionale di Astrofisica (INAF).
AM and GM acknowledge support from the European Union - Next Generation EU through 
the grant n. 2022J7ZFRA - Exo-planetary Cloudy Atmospheres and Stellar High energy 
(Exo-CASH), funded by MUR--PRIN 2022, and the ASI--INAF agreement 2021-5-HH.2-2024.
AAV acknowledges funding from the European Research Council (ERC) under the 
European Union's Horizon 2020 
research and innovation programme (grant agreement No 817540, ASTROFLOW) and funding from the Dutch 
Research Council (NWO), with project number VI.C.232.041 of the Talent Programme Vici.
KV acknowledges the support of the Hungarian National Research, Development and Innovation Office (NKFIH) Élvonal grant KKP 143986.
SB acknowledges funding by the Dutch Research Council (NWO) under the project 
"Exo-space weather and contemporaneous signatures of star-planet interactions" 
(with project number OCENW.M.22.215 of the research programme "Open Competition Domain Science- M").
Based on observations obtained with XMM-Newton, an ESA science mission
with instruments and contributions directly funded by ESA Member States and NASA.
Based on observations collected at the European Southern Observatory under ESO 
programs 1102.C-0339(A), 0102.C-0525 and 0102.D-0483, 1102.C-0744, 1102.C-0958 
and 1104.C-0350.
This work made use of data from eROSITA, a joint German-Russian science mission with the 
support of the Deutsches Zentrum für Luft- und Raumfahrt (DLR). 
\end{acknowledgements}

\appendix
\section{\xmm\ data analysis} \label{xmmsect}
\begin{figure}
    \centering
    \includegraphics[width=0.99\linewidth]{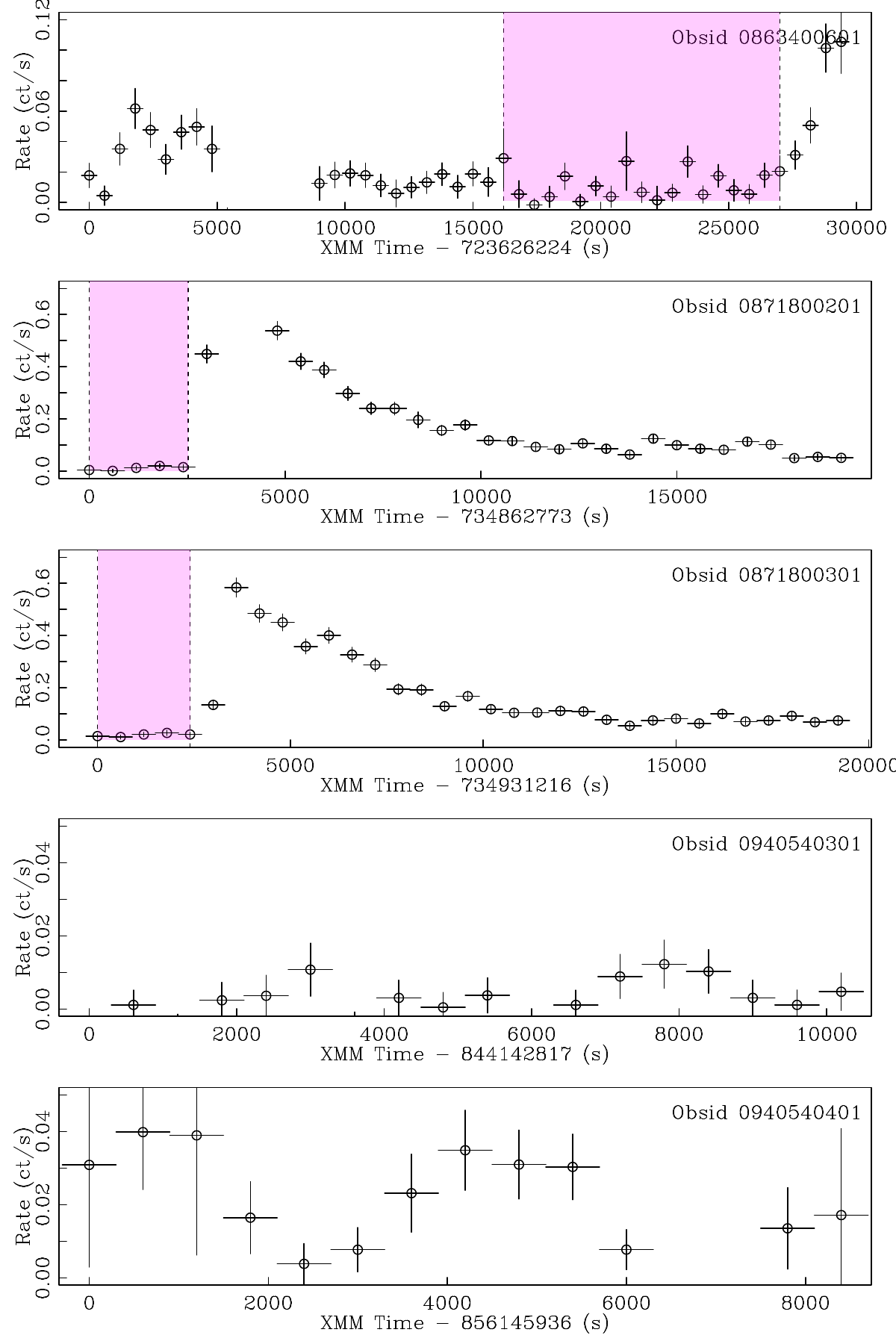}
    \caption{Light curves of \lnine\ recorded with the {\it pn} camera. 
    The shaded areas mark the intervals selected as quiescent phases in the archival observations. 
    For the observations obtained with our program (ObsID 0940540301 and 0940540401) we used the full exposure. 
    For  observations 0871800201 and 0871800301 the Y-axis scale is increased to show the entire flares.}
    \label{fig:lc}
\end{figure}

We observed \lnine\ with \xmm\ on October 1st 2024 and on February 16th 2025 
with EPIC as prime instrument (see Table \ref{xmmlog} for the details).
The files constituting the observation were downloaded and reduced with SAS 21 and SAS 22 
to obtain tables of the events recorded by the EPIC cameras MOS and {\it pn}. 
% \LEt{***abbreviation to introduce} 
The background rate was low during the first observation and it did not require further 
time filtering. In contrast, background flaring, due to space weather conditions, 
was severe during the second exposure
and thus we filtered out the high background intervals.
We retained the events in the band $0.3-10$ keV, with {\sc flag == 0} and {\sc pattern <=12}. 
Due to the close distance and the high proper motions of the star, 
the actual sky coordinates of \lnine\ and the centroid of the X-ray emission are offset with 
respect  to the J2000 coordinates by  about 8$\arcsec$. 
We also acquired images with the OM and the filter UVM2, however the star was not detected in 
this band ($200-300$ nm).
{Light curves were produced with SAS tasks {\it evselect} and {\it epiclcorr} and they 
are shown in Fig. \ref{fig:lc}.}

\begin{table*}[t]
\caption{Log of the \xmm\ observations \label{xmmlog} of \lnine.}
\begin{center}
\resizebox{0.95\textwidth}{!}{
\begin{tabular}{ccccccc}\\ \hline \hline
Observation Identifier   &     R.A. (J2000) &   Dec (J2000) & XMM orbit &  Start Date  (UT) & P.I. & Time (ks) \\\hline
863400601	 & 	08 18 07.61	 & 	-68 18 46.8	 & 	3845	 & 	2020-12-06 06:27:06.000	 & 	Wolk & 8.0  \\
871800201	 & 	08 18 07.98	 & 	-68 18 53.9	 & 	3910	 & 	2021-04-15 07:42:55.000	 & 	France  & 2.7  \\
871800301	 & 	08 18 07.98	 & 	-68 18 53.9	 & 	3910	 & 	2021-04-16 02:43:38.000	 & 	France  & 2.3 \\
940540301	 & 	08 18 07.61	 & 	-68 18 46.8	 & 	4544	 & 	2024-10-01 03:16:59.000	 & 	Pillitteri  & 9.7 \\
940540401	 & 	08 18 08.07	 & 	-68 18 55.6	 & 	4614	 & 	2025-02-16 22:19:03.000	 & 	Pillitteri  & 4.7 \\\hline
\end{tabular} 
}
\end{center}
\tablefoot{The last column lists the quiescent time  or the good time intervals (GTI) recorded by the {\em pn} instrument.}
\end{table*}
At a visual inspection, the star was barely noticeable in the observation of October 2024
while it was brighter in February 2025 and in the observations made in 2020 and 2021
(Fig. \ref{fig:rate-comp}).
We performed a source detection process with a wavelet convolution based code tailored for \xmm\ 
\citep{Damiani1997b,Damiani1997a} in order to estimate the  count rate the band $0.3-10$ keV.
The code performs a wavelet convolution and find the peaks of the wavelet convolved image 
from which it derives the count rate of the sources in the initial rate image. 
A feature of the code is the ability to sum
several EPIC images from MOS and {\it pn} within an offset of about 7$\arcmin$ in order to 
get the deepest sensitivity. To do so, exposure maps of each MOS and {\it pn} 
at a resolution of 2\arcsec were built with SAS and values of the effective area weighted 
for a thermal spectrum  were used to properly scale the counts of each EPIC instrument.
We analyzed the sum of the rate images properly scaled for the effective area of each EPIC camera.
The code output consists of a list of detected sources with positions, count rates and 
detection significance, a combined event list from the input files, 
a rate image, and background and exposure maps. 
The resulting count rates are listed in Table \ref{tab:rate_lx}. 

For the archival observations we selected the quiescent emission in the 
pre-flare intervals for the observations by France (2030 s for Obsid 0871800201 
and 2400s for 0871800301),  and the middle interval for the observation by  
Wolk. 
Figure \ref{fig:lc} shows the intervals used for the quiescent emission.
We analyzed the resulting images with the same wavelet convolution procedure 
to obtain the rates and the detection significance listed in Table \ref{tab:rate_lx}.
The final count rates are scaled to the MOS instrument sensitivity.
From count rates we obtained fluxes using the PIMMS software. 
We adopted a 1T thermal APEC model with plasma at log T = 6.3 or kT = 0.17 keV, 
solar abundances and $N_H = 10^{18}$ \nhu. The choice is consistent with the 
best fit of the quiescent spectrum acquired in 2021 \citep{Behr2023}.
From fluxes we obtained luminosity values using the distance of 10.6 pc to the star.
From {\it pn} events we derived also a hardness ratio $HR=(H-S)/(H+S)$ by selecting events in 
$0.3-0.7$ keV for the soft band and 0.7-8.0 keV for the hard band. The values
for each observation are listed in Table \ref{tab:rate_lx}.

As a further check on the variability of the star in X-rays, we performed the source detection
with SAS ver. 22 with the {\it edetect\_chain} task which operates a detection based
on the smoothed image with a {\it boxcar} function and a maximum likelihood (ML) estimate 
of detection on the final list of detected sources. 
This task however does not allow to combine images from different observations as we did
with the wavelet convolution code. 
We report the ML values in Table \ref{tab:rate_lx}.
The values of ML quantify the detection significance, for \lnine\ they vary in a range between 
11.4 and  110.5 among the observations, with the lowest significance due to the faintest 
emission measured in October 2024.

\section{Optical spectra}
\label{harpssect}
% \LEt{***Additional optical spectra ? }
Optical spectra of \lnine\ were acquired between October 2018 and April 2019 
with ESO HARPS  \citep{Cloutier2019}. 
These spectra encompass the region of CaII H\&K lines at 3933~\AA\ and 3968~\AA\
allowing to calculate the S index \citep{Vaughan1978} and the \logrhk\ index 
\citep{Noyes1984}. \citet{Cloutier2019} report
the values of S index derived from the same spectra and that we used to derive 
\logrhk\ with the 
calibration for M dwarfs given by \citet{Astudillo2017}.

Spectra collected with ESPRESSO spectrograph were acquired between November 2018 and 
March 2020 allowing a partial overlap with HARPS spectra. 
We used the data obtained with the ESPRESSO DRS pipeline version 3.2.5 to obtain the S and \logrhk\ 
indexes. 
The $5\%-95\%$ quantile range of \logrhk\ { is from $-5.6$  to $-5.2$} with a median value of $-5.4$. 
A Lomb-Scargle periodogram of the series of the \logrhk\ index shows a prominent peak at 
$\sim100$ days which likely arises from an alias of the seasonal sampling window and the 
rotational period of the star (80 days). However, there is no evidence of a long-term cycle in the
chromospheric activity indicators. 

\end{document}